\documentclass[aps,prb,twocolumn,showpacs,amsmath,twocolumn,amssymb,superscriptaddress,letterpaper]{revtex4}
\usepackage{times}
\usepackage{amsfonts}
\usepackage{mathrsfs}
\usepackage{graphicx}
\usepackage{dcolumn}
\usepackage{bm}
\usepackage{color}

\usepackage[colorlinks,bookmarks=false,citecolor=blue,linkcolor=red,urlcolor=blue]{hyperref}
\begin{document}
\title{ Hybrid Crystals of Cuprates and Iron-Based Superconductors }

\author{Xia Dai}
\affiliation{Institute of Physics, Chinese Academy of Sciences, Beijing 100190, China}

\author{Congcong Le}
\affiliation{Institute of Physics, Chinese Academy of Sciences, Beijing 100190, China}

\author{Xianxin Wu}
\affiliation{Institute of Physics, Chinese Academy of Sciences, Beijing 100190, China}

\author{Jiangping Hu  }
\affiliation{Institute of Physics, Chinese Academy of Sciences, Beijing 100190, China}
\affiliation{Collaborative Innovation Center of Quantum Matter, Beijing, China}
\affiliation{Department of Physics, Purdue University, West Lafayette, Indiana 47907, USA}

\date{\today}

\begin{abstract}

We propose two possible new compounds, Ba$_2$CuO$_2$Fe$_2$As$_2$ and K$_2$CuO$_2$Fe$_2$Se$_2$, which hybridize the building blocks of two high temperature superconductors, cuprates and iron-based superconductors. These compounds consist of square CuO$_2$ layers and antifluorite-type Fe$_2$X$_2$ (X=As,Se) layers separated by Ba/K.  The calculations of binding energies and phonon spectrums indicate that they are dynamically stable, which ensures that they may be experimentally synthesized. The Fermi surfaces and electronic structures of the two compounds inherit the characteristics of both cuprates and iron-based superconductors. These  compounds can be  superconductors  with intriguing physical properties to help to determine pairing mechanisms of high $T_c$ superconductivity.

\end{abstract}


\maketitle

\section{introduction}

 Cuprates\cite{Bednorz1986,Kishio1987,Cava1987,Hazen1987,Wu1987,Maeda1988,Shengzz1988,Schilling1993}, first discovered in 1986, and iron-based superconductors\cite{Joshua2008,Marianne2008,Yoichi2008,Hsu2008,Guo2010}(IBS), discovered in early 2008, are two classes of unconventional high temperature superconductors who share many common features. Both of them are quasi-two-dimensional and the phase diagrams are similar in which superconductivity develops after a magnetic order is suppressed\cite{Andrea2003,Basov2011,Jphu2012}. All the cuprates share a common structure element CuO$_2$ plane, where Cu atoms form a square lattice\cite{Barisic2013}. The IBS share a common Fe$_2$X$_2$ (X=As and Se) layered structure unit, which possesses an anti-PbO-type (anti-litharge-type) atom arrangement. The Fe$_2$X$_2$ layers consist of a square lattice sheet of Fe coordinated by X above and below the plane to form face sharing FeX$_4$ tetrahedra\cite{Paglione2010}.

One of major questions in the field of high $T_c$ superconductors is whether cuprates and IBS share a common  superconducting mechanism\cite{Jphu2012,Scalapino2012}.  The answer to this question may be obtained if  we can integrate the characteristics of both superconductors into a single compound so that their relations can be exclusively addressed.  As both structures are featured with layered square lattices with similar in-plane lattice constants,  it is possible to design   a compound  containing both building blocks, Cu-O layers of cuprates and  Fe-As(Se) layers of IBS. A similar material design has been adopted and a target material Nd$_4$CuO$_6$Fe$_2$As$_2$ has been theoretically investigated\cite{Ricci2010}.

Recently, two new materials Ba$_2$MO$_2$Ag$_2$Se$_2$ (M = Co,Mn)\cite{Zhou2014} have been synthesized via solid-state reaction in experiment. These two compounds, whose structures belong to I4/mmm space group, consist of infinite MO$_2$ square planes and antifluorite-type Ag$_2$Se$_2$ layers separated by barium.  In this paper, motivated by the fact that the MO$_2$ plane resembles the CuO$_2$ plane in cuprates and the Ag$_2$Se$_2$ layer resembles the Fe$_2$X$_2$ layer in IBS, we consider the substitution of Cu and Fe for M and Ag  in the compounds respectively to obtain two possible new compounds, Ba$_2$CuO$_2$Fe$_2$As$_2$ and K$_2$CuO$_2$Fe$_2$Se$_2$.  We perform density functional calculations to study the stability and basic electronic structures of these new compounds.  We find that these materials integrate basic electronic characteristics of both high $T_c$ superconductors.  In Ba$_2$CuO$_2$Fe$_2$As$_2$,  the CuO$_2$ layers are  electron-doped while the Fe$_2$As$_2$  layers are hole-doped. The situation is reversed   in  K$_2$CuO$_2$Fe$_2$Se$_2$. Such doping configurations suggest that many possible mixed phases, in particular, magnetic and superconducting phases,  may be realized in these materials by introducing additional carriers and applying external pressure.

\section{Computational Details}\label{s1}
Our calculations are performed using density functional theory (DFT) employing the projector augmented wave (PAW) method encoded in the Vienna ab initio simulation package (VASP) \cite{Kresse1993,Kresse1996,Kresse1996b}. Both of the local density approximation (LDA) and generalized-gradient approximation (GGA)\cite{Perdew1996} for the exchange correlation functional are used. Throughout the work, the cutoff energy is set to be 500 eV for expanding the wave functions into plane-wave basis. In the calculations of magnetic properties, the LDA+U method is used with the effective on-site Coulomb U being 7 eV for Cu $3d$ states\cite{Czyzyk}. In the calculations, the Brillouin zone is sampled in the $\textbf{k}$ space within Monkhorst-Pcak scheme\cite{Monkhorst1976}. The number of these $\textit{k}$ points are depending on the lattice: ${15 }\times{ 15 }\times{ 3}$ for the general unit cell with 4 Fe atoms and 2 Cu atoms and ${9 }\times{ 9 }\times{ 3}$ for the $\sqrt{2} \times \sqrt{2}$ unit cell. We relax the lattice constants and internal atomic positions with both LDA and GGA, where forces are minimized to less than 0.01 eV/\AA. The phonon dispersions are calculated using finite displacement method\cite{Alfe2009} as implemented in the PHONOPY code\cite{Togo2008,Togo2015}.

\section{Crystal Structure}\label{s2}
\begin{figure}
\centerline{\includegraphics[width=0.3\textwidth]{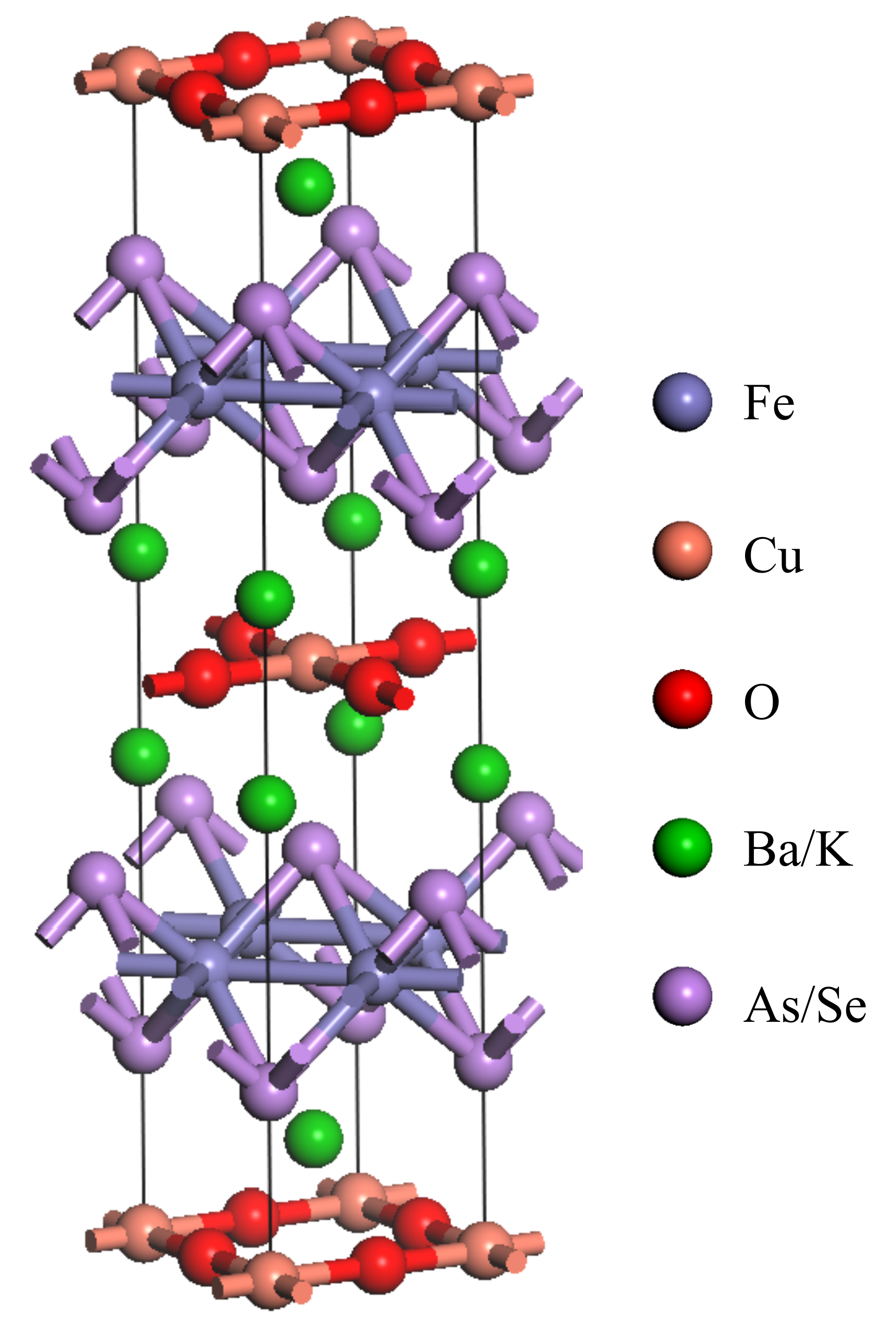}} \caption{(color online) Crystal structure of Ba$_2$CuO$_2$Fe$_2$As$_2$ (K$_2$CuO$_2$Fe$_2$Se$_2$).
 \label{structure} }
\end{figure}

 Ba$_2$CuO$_2$Fe$_2$As$_2$ and K$_2$CuO$_2$Fe$_2$Se$_2$ crystallize in a body-centered tetragonal lattice, shown in Figure \ref{structure}. The Ba (K) spacer layer separates the tetrahedra6l Fe$_2$As$_2$ (Fe$_2$Se$_2$) layers and the square CuO$_2$ layers. There are double Fe$_2$X$_2$ (X=As and Se) and CuO$_2$ layers in a unit cell, similar to BaFe$_2$As$_2$. In order to predict the structures of Ba$_2$CuO$_2$Fe$_2$As$_2$ and K$_2$CuO$_2$Fe$_2$Se$_2$, the lattices are fully optimized based on the experimental structural parameters of Ba$_2$CoO$_2$Ag$_2$Se$_2$. The optimized and experimental structural parameters are summarized in Table \ref{tab_parameter}. Both of the calculated lattice constants using LDA and GGA are close to those of Ba$_2$CoO$_2$Ag$_2$Se$_2$ in experiment, which validates the adopted substitution. As the calculated lattice parameters for Ba$_2$CoO$_2$Ag$_2$Se$_2$  using GGA are closer to the experimental data, we perform the following calculations for the two materials using the lattice parameters obtained by relaxation with GGA.

\begin{figure}
\centerline{\includegraphics[width=0.4\textwidth]{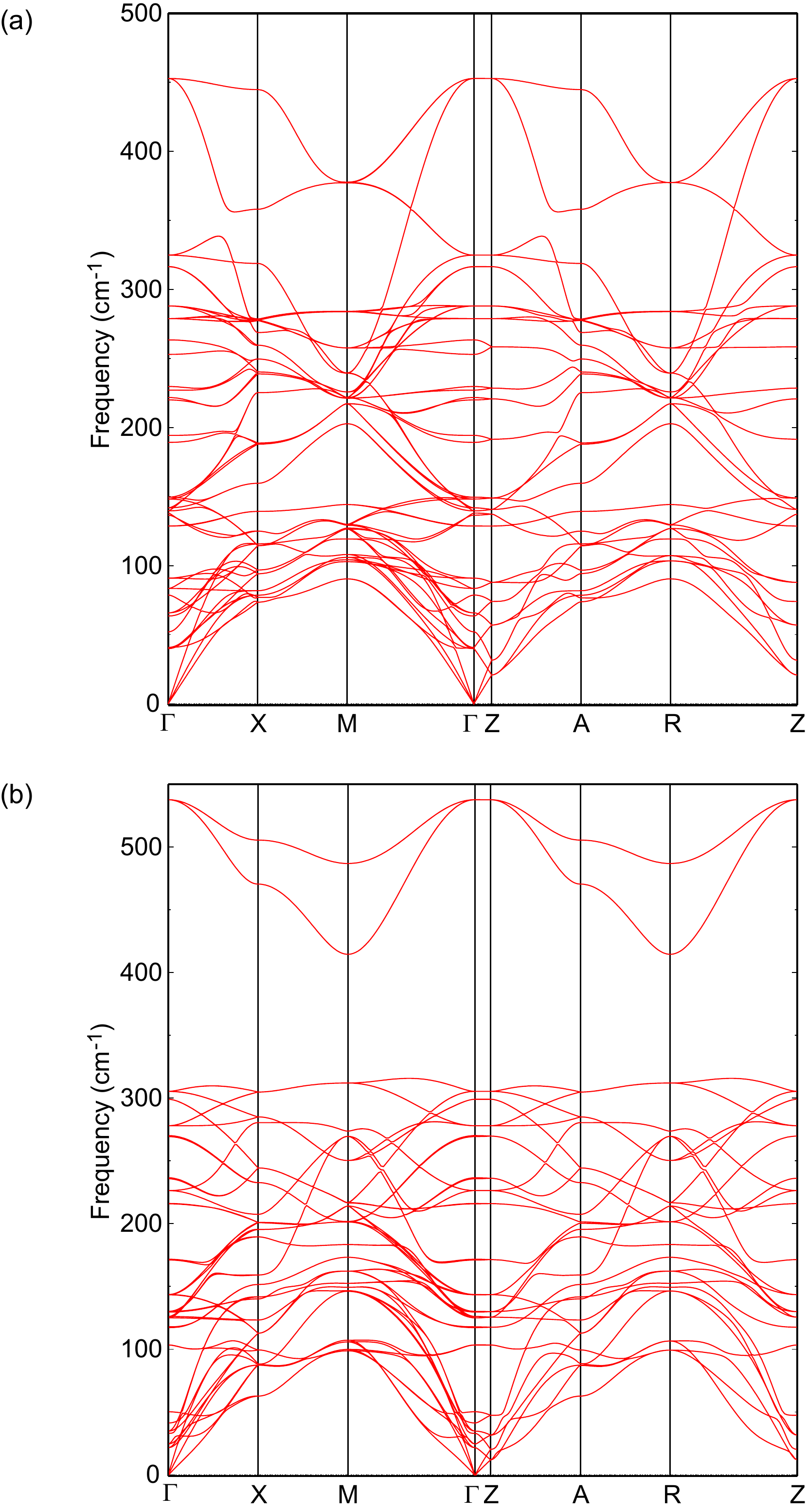}} \caption{(color online) Calculated GGA phonon dispersions of Ba$_2$CuO$_2$Fe$_2$As$_2$ and K$_2$CuO$_2$Fe$_2$Se$_2$ in the paramagnetic state.
 \label{phonon_dispersion} }
\end{figure}

The binding energy is usually calculated to estimate the stability of new structures. Here, the binding energy per atom, $E_b$, is defined as $E_b=(4E_{Ba/K}+2E_{Cu}+4E_{O}+4E_{Fe}+4E_{As/Se}-E_{total})/18$, in which $E_{Ba}$,$E_K$,$E_{Cu}$,$E_O$,$E_{Fe}$,$E_{As}$ and $E_{Se}$ are the respective energies per atom of elemental Ba, K, Cu, O, Fe, As and Se in the states at standard ambient temperature and pressure. $E_{total}$ is the calculated total energy of a unit cell of Ba$_2$CuO$_2$Fe$_2$As$_2$ (K$_2$CuO$_2$Fe$_2$Se$_2$). The obtained binding energies are 1.61 eV and 1.23 eV per atom for Ba$_2$CuO$_2$Fe$_2$As$_2$ and K$_2$CuO$_2$Fe$_2$Se$_2$, respectively. Both of the binding energies are very close to that of Ba$_2$CoO$_2$Ag$_2$Se$_2$(1.318 eV), indicating that the two structures are energetically favorable in experiment and may be synthesized using similar methods. To further test the stability of these two structures, we calculate their phonon dispersions, shown in Figure \ref{phonon_dispersion}. No imaginary frequencies are observed throughout the whole Brillouin zone in phonon dispersions, confirming their dynamically structural stability.

\begin{table}
\caption{Optimized structural parameters of Ba$_2$CuO$_2$Fe$_2$As$_2$ and K$_2$CuO$_2$Fe$_2$Se$_2$, using LDA and GGA in the paramagnetic phase. The 6th column is the experimental structural parameters of Ba$_2$CoO$_2$Ag$_2$Se$_2$\cite{Zhou2014}.}
\label{tab_parameter}
\begin{ruledtabular}
\begin{tabular}{c|cc|cc|c}
&\multicolumn{2}{c|}{Ba$_2$CuO$_2$Fe$_2$As$_2$} &\multicolumn{2}{c|}{K$_2$CuO$_2$Fe$_2$Se$_2$} & Ba$_2$CoO$_2$Ag$_2$Se$_2$ \\
\colrule
&LDA&GGA&LDA&GGA&Expt.\\
\colrule
$a$(\AA) &  3.902 &  4.026 & 3.769  & 3.839  &  4.223 \\
$c$(\AA) & 19.867 & 20.273 & 18.320 & 22.165 & 20.036 \\
\end{tabular}
\end{ruledtabular}
\end{table}

\begin{table}
\caption{Optimized structural parameters of Ba$_2$CuO$_2$Fe$_2$As$_2$ using GGA in the paramagnetic phases and those of LiFeAs\cite{Joshua2008}, BaFe$_2$As$_2$\cite{Marianne2008} and LaFeAsO\cite{Yoichi2008}. The angle $\beta$ is the Fe-As-Fe bonding angle of two next nearest Fe atoms.}
\label{tab_Ba}
\begin{ruledtabular}
\begin{tabular}{ccccc}
& Ba$_2$CuO$_2$Fe$_2$As$_2$ & LiFeAs & BaFe$_2$As$_2$ & LaFeAsO \\
\colrule
$a$(\AA) & 4.026 & 3.791 & 3.963 & 4.035 \\
$\beta$($\circ$) & 118.7 & 103.1 & 111.0 & 113.5 \\
h(Fe-As)(\AA) & 1.19 & 1.51 & 1.36 & 1.32 \\
h(Fe-Fe)(\AA) & 10.14 & 6.36 & 6.51 & 8.74 \\
\end{tabular}
\end{ruledtabular}
\end{table}

\section{Electronic and Magnetic Properties}\label{s3}
The optimized parameter $a$ of Ba$_2$CuO$_2$Fe$_2$As$_2$ is determined to be 4.026\text{\AA}, slightly bigger than those of cuprates (3.78\text{\AA} for La$_{1.8}$Sr$_{0.2}$CuO$_4$\cite{Cava1987} and 3.859\text{\AA} for YBa$_2$Cu$_3$O$_{6+x}$\cite{Hazen1987}). Compared with three typical kind of iron pnictides, LiFeAs\cite{Joshua2008}, BaFe$_2$As$_2$\cite{Marianne2008}, and LaFeAsO\cite{Yoichi2008}, as shown in Table \ref{tab_Ba}, the parameter $a$ is very close to those of BaFe$_2$As$_2$ and LaFeAsO but deviates a little from that of LiFeAs. The obtained As height above the Fe plane in Ba$_2$CuO$_2$Fe$_2$As$_2$ is smaller than those of iron pnictides in experiment. This underestimation has been noted in the study of IBS\cite{Yin2008,Mazin2008a,Singh2008b}. The Fe$_2$As$_2$ inter-layer distance is much larger than the values of the three families in iron pnictides, indicating that it is more two-dimensional than conventional iron pnictides. The separation between Fe$_2$As$_2$ and CuO$_2$ layers is about 5\AA,~ which suggests couplings between these layers.

\begin{figure}
\centerline{\includegraphics[width=0.4\textwidth]{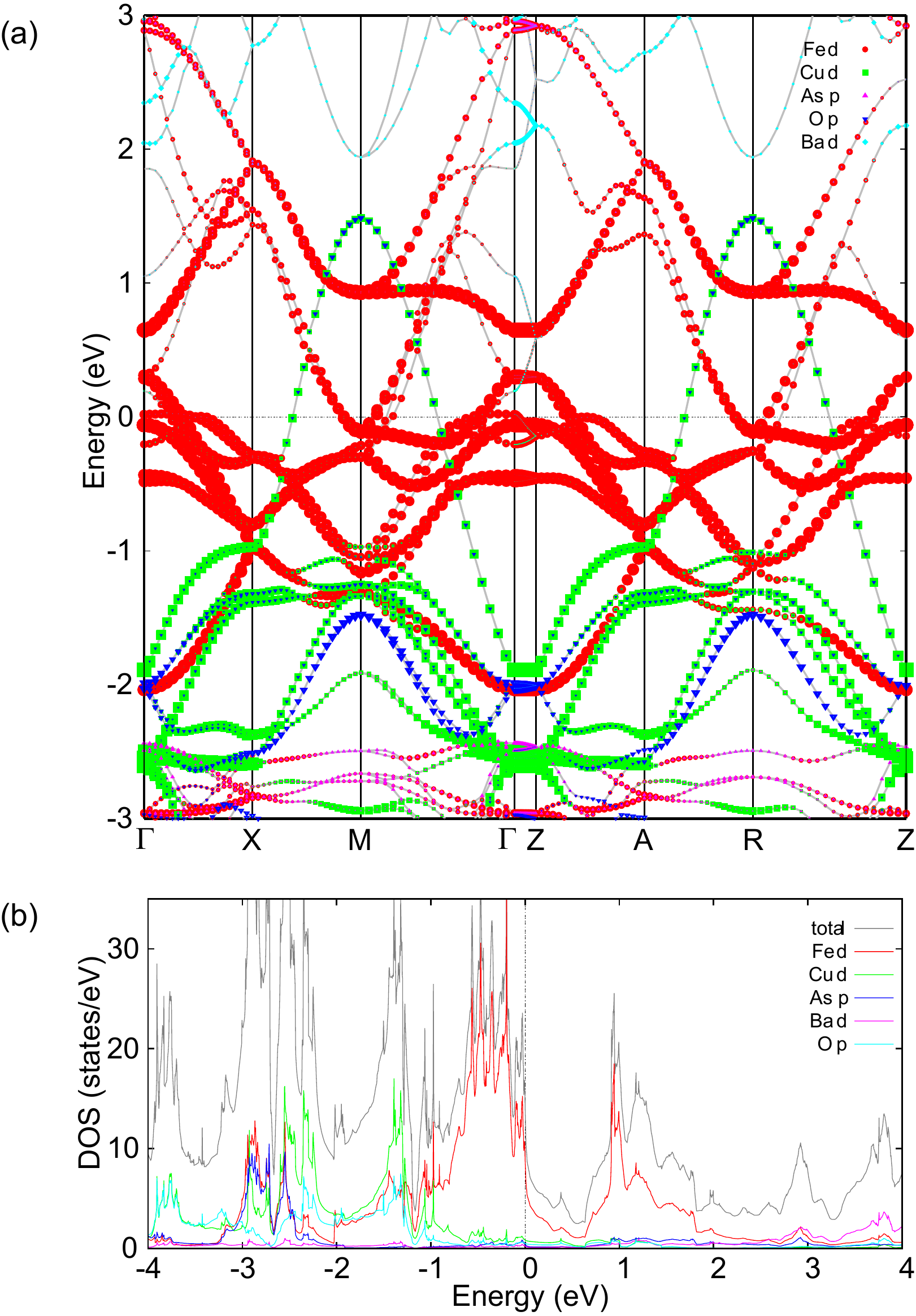}} \caption{(color online) Band structures and projected density of states of Ba$_2$CuO$_2$Fe$_2$As$_2$ using GGA relaxed parameters in the paramagnetic state.
 \label{Ba_pm} }
\end{figure}

The band structure and density of states (DOS) for Ba$_2$CuO$_2$Fe$_2$As$_2$ with GGA optimized structural parameters in the paramagnetic states are shown in Figure \ref{Ba_pm}. The band structure near the Fermi level resembles those of both typical iron pnictides LaFeAsO\cite{Singh2008} and cuprates\cite{Andrea2003} where Fe 3d states ($d_{xz},d_{yz},d_{x^2-y^2}$ orbitals) and Cu 3d states ($d_{x^2-y^2}$ orbitals) dominate the Fermi level.
Similar to iron pnictides, it is a bad metal with low carrier density. The 3$d$ states of Fe are mainly located near the Fermi level from -2.5 eV to 3.0 eV and a pseudogap appears at an electron count of six. The As $p$ states mainly lie 2.5 eV below the Fermi level but slightly mix with the Fe 3$d$ states in the energy range from -1.0 eV to 1.0 eV. The Cu 3$d$ contribution is concentrated between -3.0 eV and 1.5 eV. The O $p$ states are strongly coupled with 3$d$ states of Cu in the total energy range. The spacer layer Ba 5$d$ and 6$s$ states, which donate electrons to Fe$_2$As$_2$ and CuO$_2$ layers, are empty and lie 2 eV above the Fermi level. At the Fermi level, the total DOS shows the same negative slope with the minimum slightly above the Fermi level, similar to the conventional IBS within DFT. The DOS at the Fermi energy is about 7.81 eV$^{-1}$/f.u. for both spins. Considering that there are two Fe atoms in one formula unit for Ba$_2$CuO$_2$Fe$_2$As$_2$ but only one for LaFeAsO, this value (3.91 eV$^{-1}$/Fe)  is larger than that of LaFeAsO (2.62 eV$^{-1}$/f.u. \cite{Singh2008}). The corresponding Pauli susceptibility and specific heat coefficient are $\chi_0=2.53 \times 10^{-4}$ emu/mol and $\gamma=18.4$ mJ/($K^2$ mol).

\begin{figure}
\centerline{\includegraphics[width=0.3\textwidth]{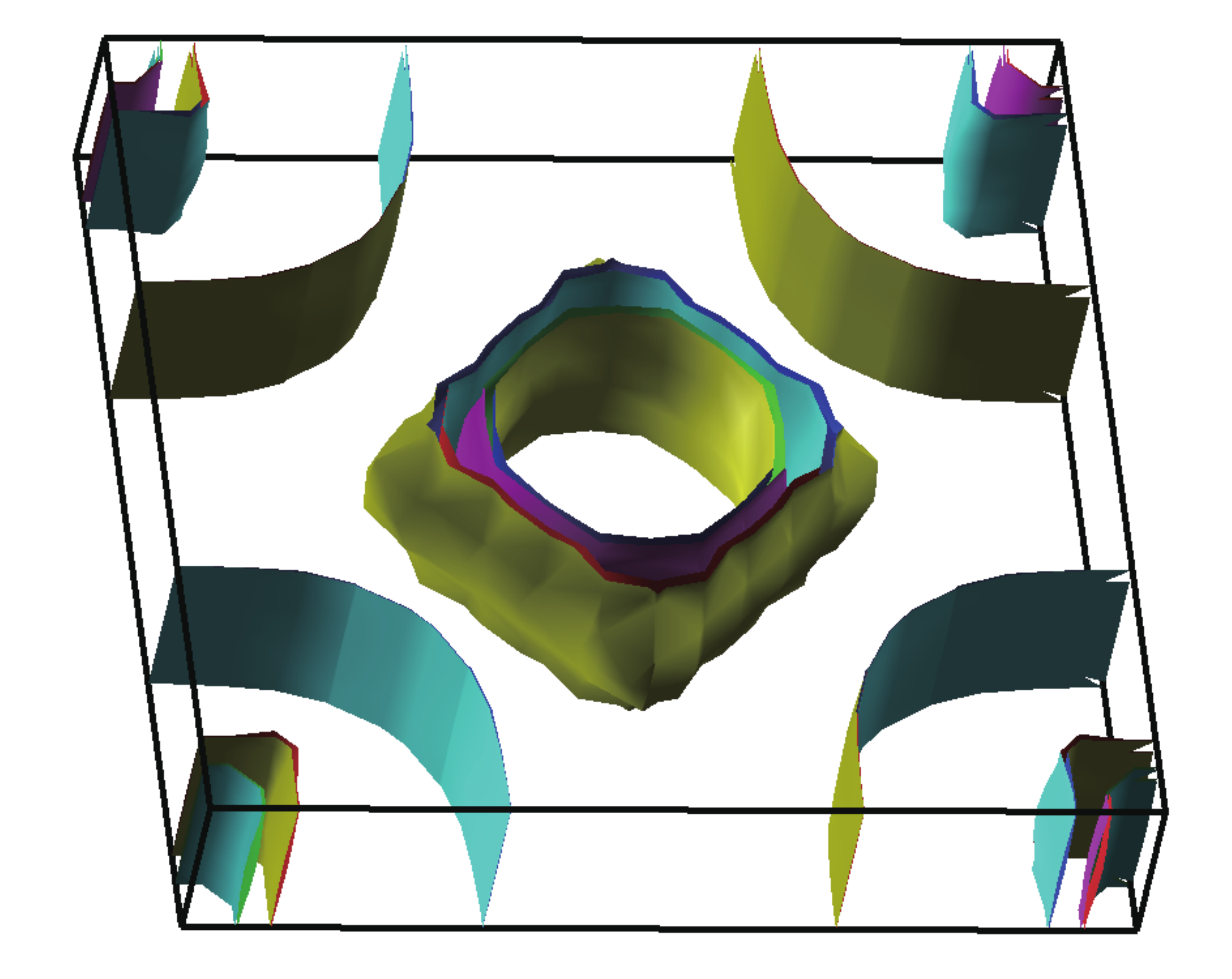}} \caption{(color online) Calculated Fermi surfaces of Ba$_2$CuO$_2$Fe$_2$As$_2$ using GGA relaxed parameters in the paramagnetic state.
 \label{fs_Ba_pm} }
\end{figure}

The calculated Fermi surfaces of Ba$_2$CuO$_2$Fe$_2$As$_2$ are given in Figure \ref{fs_Ba_pm}, which are very similar to those of both iron pnictides and cuprates. From Figure \ref{Ba_pm}(a), we find that the Fe $d_{xz}$ and $d_{yz}$ states yield two hole cylinders at the zone center. There is an additional heavy 3D hole pocket centered at $\Gamma$ point, which intersects and anticrosses with the hole cylinders. The 3D pocket is derived from Fe $d_{z^2}$ states which hybridize with As $p_z$ states. At the zone corner, there are two 2D small electron pockets and one 2D large hole pocket. The electron pockets are mainly attributed to Fe $d_{xz}$, $d_{yz}$ and $d_{x^2-y^2}$ orbitals ($d_{xy}$ orbitals in usual Fe lattice). The hole pocket around M point, however, is mainly derived from Cu $d_{x^2-y^2}$ and O $p_x$ and $p_y$ states, in accordance with the Fermi surfaces of cuprates\cite{Andrea2003}.
All of the Fermi surface sheets are double degenerate as there are two Fe$_2$As$_2$ and CuO$_2$ layers in one unit cell.

It is interesting to see that the hole pockets at $\Gamma$ point are large, which indicates heavy hole doping in the Fe$_2$As$_2$ layers.  From the size of the pockets, we can estimate that the hole doping concentration with respect to BaFe$_2$As$_2$ is roughly 0.15 holes per Fe atom. Considering the hole doping level, the Fermi surfaces attributed to Fe$_2$As$_2$ layers should be similar to Ba$_{0.7}$K$_{0.3}$Fe$_2$As$_2$ except that the pockets here are much more two-dimensional. This hole doping level is very close to that in optimally-hole-doped pnictide Ba$_{0.6}$K$_{0.4}$Fe$_2$As$_2$\cite{Ding2008,Zhao2008,Nakayama2009} where $T_c$ can reach 37 K.  Actually, the calculated Fermi surfaces are similar to those of (Ba,K)Fe$_2$As$_2$ obtained in ARPES experiment\cite{ChangLiu2008,Xu2011}. As required by the charge conservation, the heavy hole doping in Fe$_2$As$_2$ layers suggests that CuO$_2$ layers must be heavily electron-doped with the same doping concentration. This consistency is confirmed by the estimation that the size of the largest hole pocket at M point attributed to $d_{x^2-y^2}$ orbitals in CuO$_2$ layers is about 0.3 electrons per Cu atom from their half filling configuration. The doping level is a little far from narrow superconducting region for electron-doped cuprates in the phase diagram\cite{Andrea2003}.

\begin{figure}
\centerline{\includegraphics[width=0.4\textwidth]{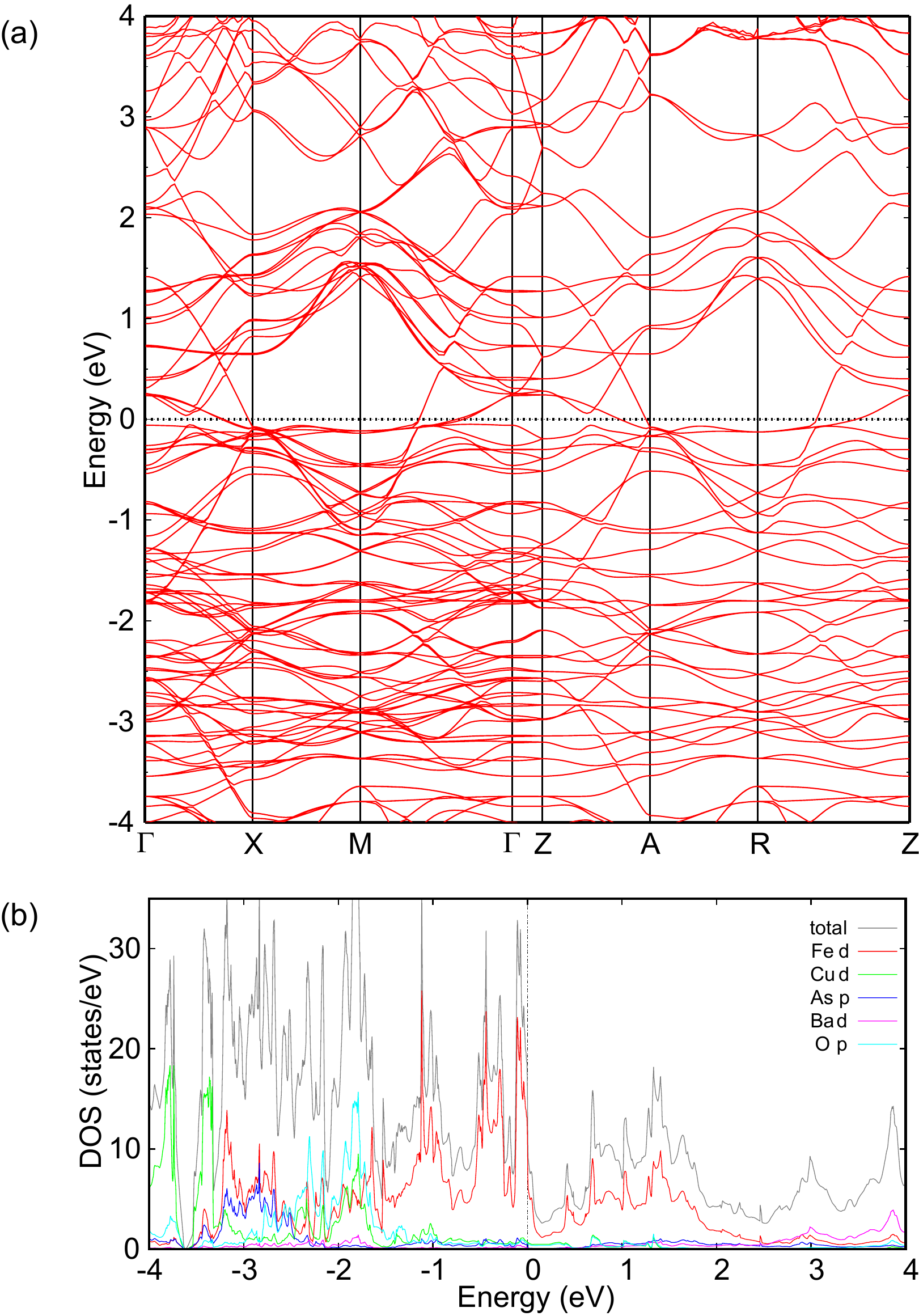}} \caption{(color online) Band structures and projected density of states of Ba$_2$CuO$_2$Fe$_2$As$_2$ using GGA relaxed parameters in the E-type collinear AFM state.
 \label{Ba_afm} }
\end{figure}

The superconductivity in both cuprates and IBS is mostly likely related to the magnetism. Therefore, we investigate the magnetic properties for this new compound. With such a large doping in the Fe$_2$As$_2$ and CuO$_2$ layers,  it is expected that the  checkboard  antiferromagnetic (AFM) order in the CuO$_2$ layers  and the E-type collinear AFM order in Fe$_2$As$_2$ layers should be significantly weakened or completely suppressed even in the meanfield type of DFT calculations. To check this, we consider four possible magnetic states for Fe$_2$As$_2$ and CuO$_2$ layers: paramagnetic state, ferromagnetic state, checkerboard AFM state and collinear AFM state. We calculated the total energies of these magnetic states using the LDA+U approach. The calculations show that there is no statically ordered moment in CuO$_2$ layers  while  the E-type collinear AFM order in Fe$_2$As$_2$ layers is still possible.  We find that in the DFT result, the E-type state has an energy gain of 103.5 meV/f.u. relative to the paramagnetic state and a spin moment of 1.59 $\mu_B$ for each Fe.  It is known that magnetic moments calculated on the known parent compounds of IBS, for example, BaFe$_2$As$_2$, is  typically over 2.0$\mu_B$\cite{Singh2008b,Akturk2008}. This moment value is significantly lowered, which is consistent with our expectation as Fe$_2$As$_2$ layers are overdoped by holes.  Figure \ref{Ba_afm} (a) and (b) show the calculated band structure and DOS in the E-type collinear AFM state (with $\sqrt{2} \times \sqrt{2}$ supercell) with internal coordinates fixed to the values obtained by non-spin-polarized energy minimization. The E-type state is still metallic with low carrier concentration and N($E_F$) decreases severely. Among the considered magnetic states above, the checkerboard AFM state in Fe$_2$As$_2$ layers is found to be metastable state relative to the E-type collinear AFM state. The energy gain is 81.5 meV/f.u. relative to the paramagnetic state and the magnetic moment is about 1.58 $\mu_B$ for each Fe atom.

\begin{figure}
\centerline{\includegraphics[width=0.4\textwidth]{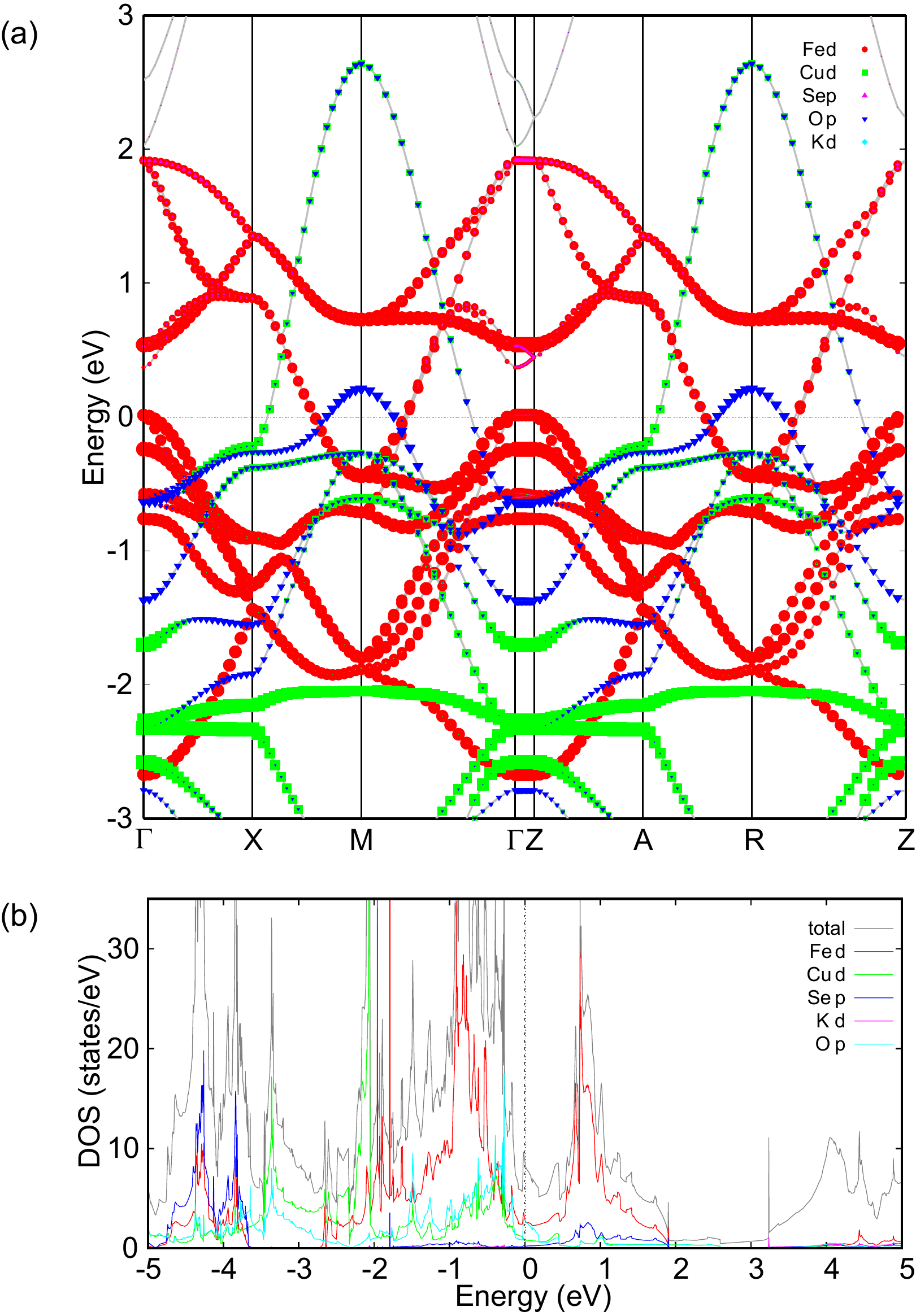}} \caption{(color online) Band structures and projected density of states of K$_2$CuO$_2$Fe$_2$Se$_2$ using GGA relaxed parameters in the paramagnetic state.
 \label{K_pm} }
\end{figure}

\begin{figure}
\centerline{\includegraphics[width=0.3\textwidth]{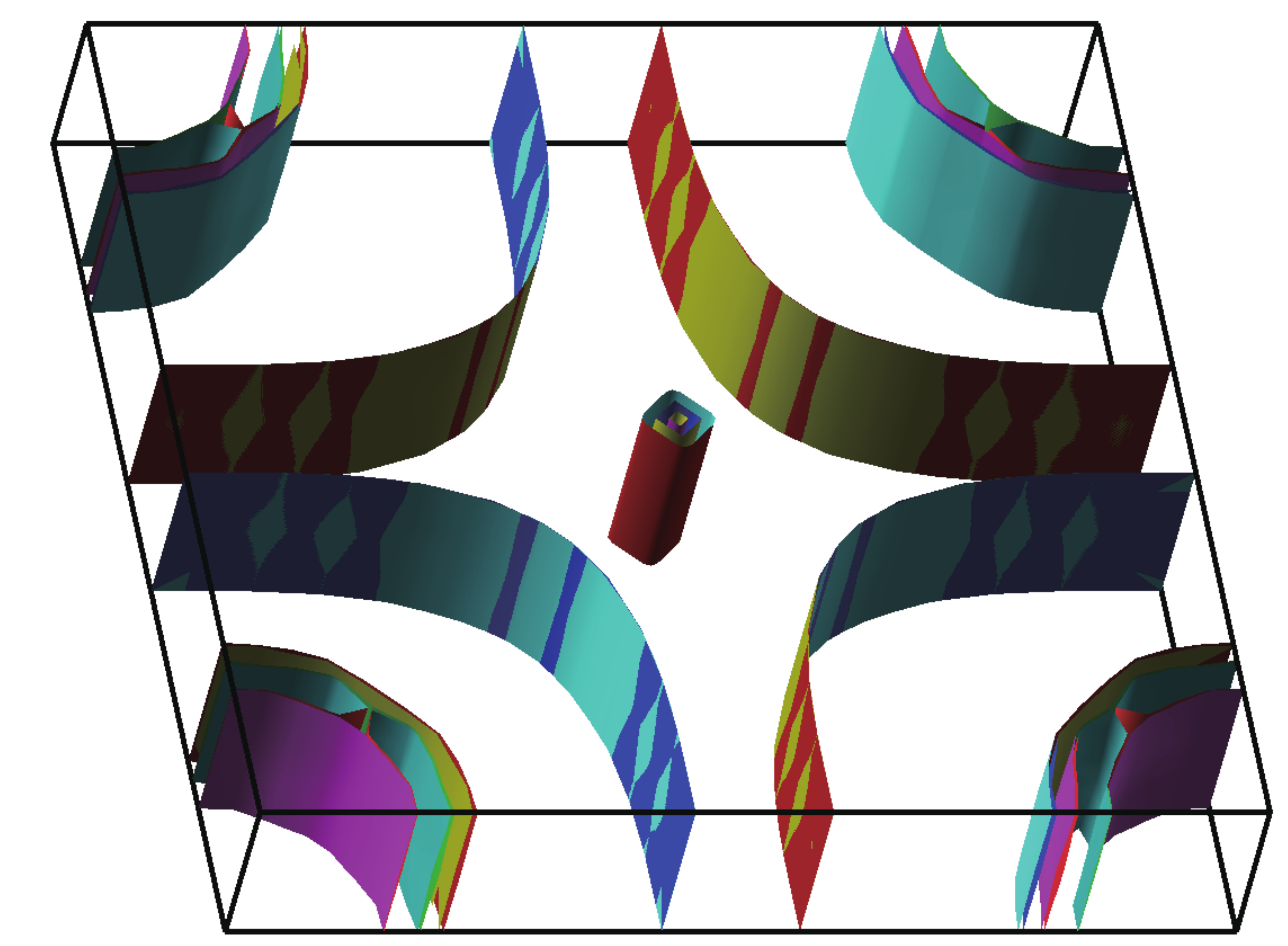}} \caption{(color online) Calculated Fermi surfaces of K$_2$CuO$_2$Fe$_2$Se$_2$ using GGA relaxed parameters in the paramagnetic state.
 \label{fs_K_pm} }
\end{figure}

With the substitution of As by Se and Ba by K in Ba$_2$CuO$_2$Fe$_2$As$_2$, we obtain a new compound K$_2$CuO$_2$Fe$_2$Se$_2$. Compared with Ba$_2$CuO$_2$Fe$_2$As$_2$, the parameter $a$ decreases to 3.839 \AA~but $c$ increases to 22.165 \AA, which is consistent with those of iron chalcogenides. Table \ref{tab_K} lists the calculated lattice parameters using GGA and those of KFe$_2$Se$_2$\cite{Guo2010} and FeSe\cite{Alaska2008}. The calculated Se height and Se-Fe-Se bond angle are very close to the values of FeSe, indicating that they may share some common electronic and magnetic properties. The Se heights are severely underestimated in the calculations compared with those in experiment\cite{Lehman2010}. The band structure and DOS are shown in Figure \ref{K_pm}. The band structure is similar to that of Ba$_2$CuO$_2$Fe$_2$As$_2$. The DOS between -2.5 eV and 2.0 eV is dominated by Fe 3$d$ states and Se 3$p$ states are mainly located 3.5 eV below the Fermi level. The Cu 3$d$ states and O 2$p$ states are strongly coupled from -4.0 eV to 2.5 eV. The K atoms have no contribution to the bands near the Fermi level. To analyze the orbital characters near the Fermi level, we plot the fat band, shown in Figure \ref{K_pm}(a). The DOS at the Fermi level is 4.54 eV$^{-1}$/f.u., which is significantly lower than that of Ba$_2$CuO$_2$Fe$_2$As$_2$. The corresponding Pauli susceptibility and specific heat coefficient are $\chi_0=1.47 \times 10^{-4}$ emu/mol and $\gamma=10.7$ mJ/($K^2$ mol). Figure \ref{fs_K_pm} shows the calculated Fermi surfaces. The Fermi surfaces are two-dimensional cylinders. The Fe 3$d$ states yield two hole pockets at $\Gamma$ point and two small electron pockets at M point. The hole pockets are attributed to $d_{xz}$ and $d_{yz}$ states while the electron pockets are derived from Fe $d_{xz}$, $d_{yz}$ and $d_{x^2-y^2}$ states. The Fermi surfaces are similar to those in KFe$_2$Se$_2$ where hole pockets are absent\cite{Zhang2011}. Besides of pockets from Fe$_2$Se$_2$ layers, there are additional two hole pockets around the Brillouin zone corners. The bigger pocket is mainly contributed from Cu $d_{x^2-y^2}$ states, which are strongly coupled with the O $p_x$ and $p_y$ states, while the smaller one is derived from O $p_x$ and $p_y$ states. All of the Fermi surface sheets are double degenerate just as those in Ba$_2$CuO$_2$Fe$_2$As$_2$.

\begin{table}
\caption{Optimized structural parameters of K$_2$CuO$_2$Fe$_2$Se$_2$ using GGA in the paramagnetic phases and those of KFe$_2$Se$_2$\cite{Guo2010} and FeSe\cite{Alaska2008}. The angle $\beta$ is the Fe-Se-Fe bonding angle of two next nearest Fe atoms.}
\label{tab_K}
\begin{ruledtabular}
\begin{tabular}{cccc}
& K$_2$CuO$_2$Fe$_2$Se$_2$ & KFe$_2$Se$_2$ & FeSe \\
\colrule
$a$(\AA) & 3.839 & 3.913 & 3.765 \\
$\beta$($\circ$) & 111.6 & 106.6 & 111.3 \\
h(Fe-Se)(\AA) & 1.31 & 1.46 & 1.29 \\
h(Fe-Fe)(\AA) & 11.08 & 7.02 & 5.52 \\
\end{tabular}
\end{ruledtabular}
\end{table}

Compared with the Fermi surfaces of Ba$_2$CuO$_2$Fe$_2$As$_2$, we find that the hole pockets at $\Gamma$ point of K$_2$CuO$_2$Fe$_2$Se$_2$ are much smaller while  the electron pockets at M point are much larger (see Figure \ref{fs_K_pm}), which indicate heavy electron doping in the Fe$_2$Se$_2$ layer.  From the size of the pockets, we can estimate that the electron doping concentration with respect to FeSe is roughly 0.16 electrons per Fe atom.  Similarly,  we can check the consistency required by the charge conservation. The size of the hole pockets at M point from  CuO$_2$ layers is about 0.32 holes for each CuO$_2$ layer from their half filling configuration.

\begin{figure}
\centerline{\includegraphics[width=0.4\textwidth]{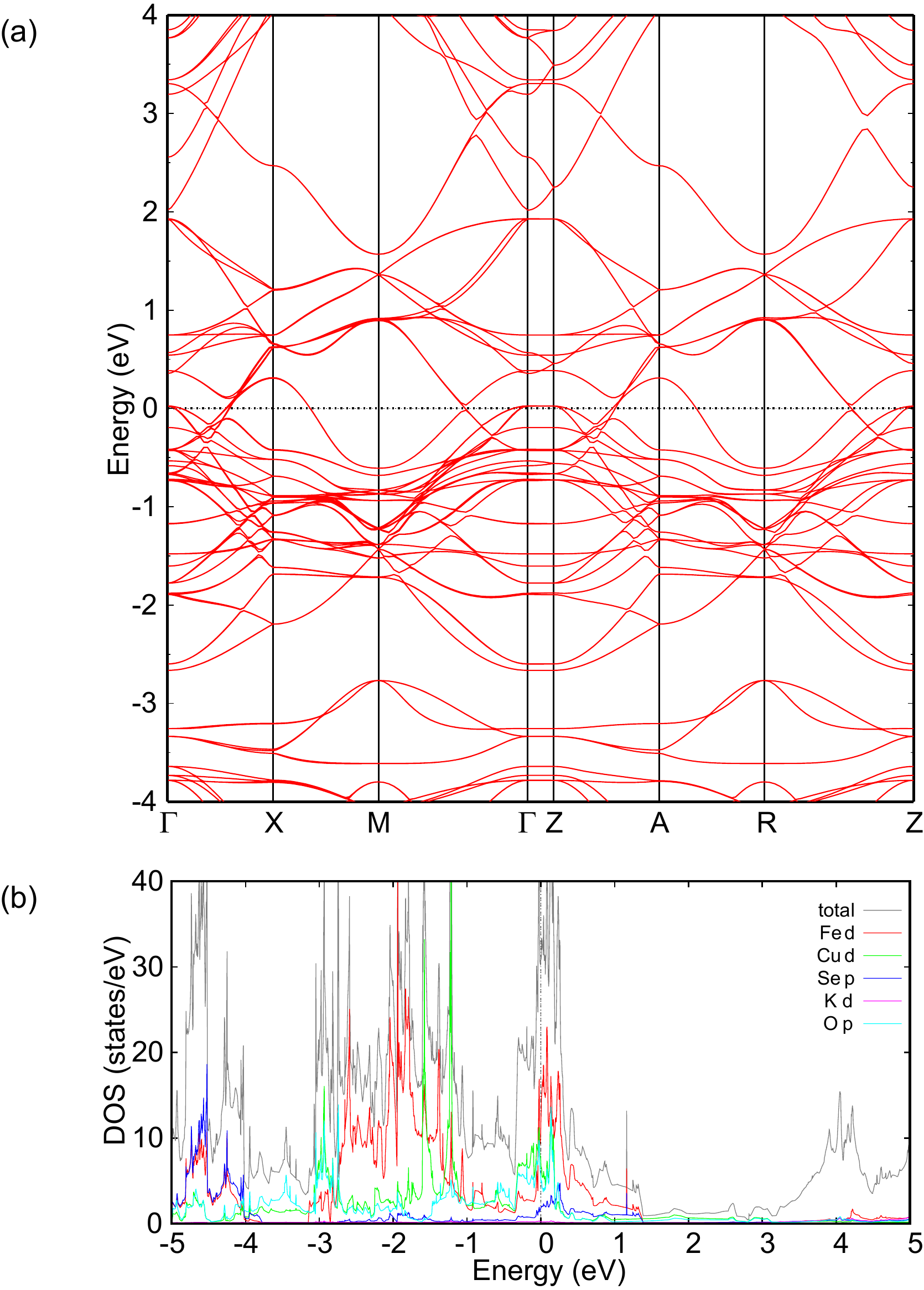}} \caption{(color online) Band structures and projected density of states of K$_2$CuO$_2$Fe$_2$Se$_2$ using GGA relaxed parameters in the AFM state.
 \label{K_afm} }
\end{figure}

The DOS at the Fermi level attributed to Fe 3d orbitals is significantly decreased with such a large electron doping, which may suggest that magnetic order in Fe$_2$Se$_2$ layers is strongly suppressed. Compared with Ba$_2$CuO$_2$Fe$_2$As$_2$, we consider an additional bicollinear magnetic state, which is the ground state in FeTe\cite{Ma2009}. The ground state is paramagnetic on Fe$_2$Se$_2$ layers and checkboard AFM ordered for CuO$_2$ layers with a energy gain of 196.0 meV/f.u. and a spin moment of 0.59 $\mu_B$ per Cu. The band structure and DOS are shown in Figure \ref{K_afm} and the system remains metallic. In this case, the Fe 3$d$, Cu 3$d$ and O 2$p$ states dominate the Fermi level. The DOS at the Fermi level is 7.04 eV$^{-1}$/f.u. for spin up channel.

\section{Discussion}\label{s4}
It is well known that the pairing symmetry is $d$-wave for the cuprates and $s$-wave  for iron pnictides  with a potential sign change between the hole and electron Fermi pockets. As the proposed new materials contain both CuO$_2$ layers and Fe$_2$X$_2$(X=As,Se) layers and the electronic structures are similar to both, it can be expected that  two different pairing symmetries  may coexist in a single material.  The proximity effect between two types of layers with different pairing symmetries can result in interesting novel phenomena. The time reversal symmetry may be broken with mixed s-wave and d-wave pairing symmetries.

Introducing additional carriers or applying external pressure  in these materials  can maximize $T_c$.  For Ba$_2$CuO$_2$Fe$_2$As$_2$, the doping of Fe$_2$As$_2$ layer is 0.15 holes per Fe and that of CuO$_2$ is 0.3 electrons per Cu. The doping for Fe$_2$As$_2$ layer is near the optimal level but that for CuO$_2$ layer is unfortunately not in the superconducting region. To realize superconductivity for both layers, we can introduce hole doping with partial replacement of Ba$^{2+}$ ions by K$^+$ ions. From the phase diagrams of cuprates and (Ba,K)Fe$_2$As$_2$, the above case can be realized in Ba$_{2-x}$K$_x$CuO$_2$Fe$_2$As$_2$ (0.24 $\leq x \leq 0.32$) by assuming that the Fe$_2$As$_2$ and CuO$_2$ layers are equally hole doped. In this doping region, the T$_c$ for Fe$_2$As$_2$ layers shows little change. $x=0.3$ corresponds the optimal doping value for electron-doped cuprates\cite{Andrea2003}. For K$_2$CuO$_2$Fe$_2$Se$_2$, however, the Fe$_2$Se$_2$ layers are electron doped and the hole doping for CuO$_2$ layers is not in superconducting region but very close to it. The superconductivity for both layers can be realized in K$_{2-x}$Ba$_x$CuO$_2$Fe$_2$Se$_2$ (0.10 $\leq x \leq 0.54$). This doping region is much wider than that in Ba$_2$CuO$_2$Fe$_2$As$_2$ due to the wide hole-doped superconducting region in cuprates. The optimal doping value for CuO$_2$ layers is $x=0.34$. The pressure is also an effective way to tune the superconductivity.

\section{Conclusion}\label{s5}
In conclusion, we identify two hypothetical compounds  Ba$_2$CuO$_2$Fe$_2$As$_2$ and K$_2$CuO$_2$Fe$_2$Se$_2$ containing Fe$_2$X$_2$ layers and CuO$_2$ layers, which are the basic structural units of IBS and cuprates, respectively. The metallic spacer Ba(K) separates the basic units. The calculations of binding energies and phonon spectrums indicate that they are dynamically stable, ensuring that they may be experimentally synthesized. The Fermi surfaces derived from Fe$_2$As$_2$(Fe$_2$Se$_2$) layers and CuO$_2$ layers are very similar to those of iron pnictides (iron chalcogenides) and cuprates, respectively. With heavy self-doping, the ground state of Ba$_2$CuO$_2$Fe$_2$As$_2$ is determined to be E-type collinear AFM state in the Fe$_2$As$_2$ layer and paramagnetic state in the CuO$_2$ layer, while K$_2$CuO$_2$Fe$_2$Se$_2$ favors checkboard AFM state in CuO$_2$ layers and paramagnetic state in Fe$_2$Se$_2$ layers. Without external doping, superconductivity can only be achieved in Fe$_2$X$_2$ layers. However, with external doping through substitution, superconductivity for both Fe$_2$X$_2$ and CuO$_2$ layers can be simultaneously achieved in Ba$_{2-x}$K$_x$CuO$_2$Fe$_2$As$_2$ (0.24 $\leq x \leq 0.32$) and K$_{2-x}$Ba$_x$CuO$_2$Fe$_2$Se$_2$ (0.10 $\leq x \leq 0.54$). The synthesis of these new compounds will not only provide us a unique opportunity to explore exotic properties in cuprates and IBS simultaneously but also help us to understand  the mechanism of high T$_c$ superconductivity.

{\it Acknowledgments:}
The work is supported by the Ministry of Science and Technology of China 973 program( No. 2015CB921300), National Science Foundation of China (Grant No. NSFC-1190020, 11334012), and the Strategic Priority Research Program of CAS (Grant No. XDB07000000).

\end{document}